\documentclass[prb,superscriptaddress,twocolumn,showpacs,amsmath,amssymb]{revtex4}
\usepackage{graphicx}
\usepackage{bm}

\begin{document}

\title {Evolution of the electronic structure from electron-doped to
hole-doped states in the two-dimensional Mott-Hubbard system
La$_{1.17-x}$Pb$_x$VS$_{3.17}$}

\author{A. Ino}
 \altaffiliation[Present address: ]{Graduate School of Science, Hiroshima University, Higashi-Hiroshima 739-8526, Japan.}
\author{T. Okane}
\author{S.-I. Fujimori}
\affiliation{Synchrotron Radiation Research Center, Japan Atomic 
Energy Research Institute, Mikazuki, Hyogo 679-5148, Japan}

\author{A. Fujimori}
\affiliation{Synchrotron Radiation Research Center, Japan Atomic 
Energy Research Institute, Mikazuki, Hyogo 679-5148, Japan}
\affiliation{Department of Complexity Science and Engineering,
University of Tokyo, Bunkyo-ku, Tokyo 113-0033, Japan}

\author{T. Mizokawa}
\affiliation{Department of Complexity Science and Engineering, University of Tokyo, 
Bunkyo-ku, Tokyo 113-0033, Japan}

\author{Y. Yasui}
\author{T. Nishikawa}
\author{M. Sato} 
\affiliation{Department of Physics, Nagoya University, Chikusa-ku, 
Nagoya 464-8602, Japan} 

\begin{abstract}
	The filling-controlled metal-insulator transition (MIT) in a
	two-dimensional Mott-Hubbard system
	La$_{1.17-x}$Pb$_x$VS$_{3.17}$ has been studied by
	photoemission spectroscopy.  With Pb substitution $x$,
	chemical potential $\mu$ abruptly jumps by $\sim 0.07$ eV
	between $x=0.15$ and 0.17, indicating that a charge gap is
	opened at $x\simeq 0.16$ in agreement with the Mott insulating
	state of the $d^2$ configuration.  When holes or electrons are
	doped into the Mott insulator of $x\simeq 0.16$, the gap is
	filled and the photoemission spectral weight at $\mu$,
	$\varrho(\mu)$, gradually increases in a similar way to the
	electronic specific heat coefficient, although the spectral
	weight remains depressed around $\mu$ compared to that
	expected for a normal metal, showing a pseudogap behavior in
	the metallic samples.  The observed behavior of $\varrho(\mu)
	\to 0$ for $x \to 0.16$ is contrasted with the usual picture
	that the electron effective mass of the Fermi-liquid system is
	enhanced towards the metal-insulator boundary.  With
	increasing temperature, the gap or the pseudogap is rapidly
	filled up, and the spectra at $T=300$ K appears to be almost
	those of a normal metal.  Near the metal-insulator boundary,
	the spectra around $\mu$ are consistent with the formation of 
	a Coulomb gap, suggesting the influence of long-range
	Coulomb interaction under the structural disorder intrinsic to
	this system.
\end{abstract}




\maketitle

\section{Introduction}


For more than a decade, extensive studies have been devoted to
clarifying how the electronic structure evolves in filling-controlled
metal-insulator transition (MIT) systems in the presence of strong
electron correlation,\cite{ImadaFujimoriTokura} stimulated by the
discovery of  high-$T_\textrm{c}$ superconductivity in the
cuprates.  Empirically, many three-dimensional perovskite-type
3$d$-transition-metal oxides become insulators when the band filling
approaches an integer number.\cite{Fujimori} For example,
La$_{1-x}$Sr$_x$TiO$_3$ shows a filling-controlled MIT at $x\simeq
0.06$; the electron effective mass $m^*$ is enhanced on approaching
the metal-insulator phase boundary from the metallic
side.\cite{Katsufuji,Kumagai,Yoshida} Simultaneously, the
quasiparticle (QP) band at the chemical potential $\mu$ becomes narrower,
and the spectral weight is transferred from the QP band to the lower
Hubbard band.\cite{Yoshida,Rozenberg}  In a two-dimensional system
La$_{2-x}$Sr$_x$CuO$_4$, on the other hand, the effective mass $m^*$
decreases towards the metal-insulator phase boundary, 
a pseudogap in the normal state is formed and eventually
evolves into the Mott insulating gap.\cite{ino98,Momono} In order to
induce the relationship between the dimensionality and the behavior of
the electronic states near MIT, it is necessary to study the
electronic structure of other two-dimensional filling-controlled
systems.  In addition, when we discuss the behavior of the
filling-controlled MIT, we should also consider the effect of
structural disorder, which is inevitably introduced into the system by
atom substitution to realize the filling control.  Under the influence of disorder
in real systems,\cite{Sarma} a small number of doped carriers are necessarily
 Anderson-localized  and thus feel a long-range Coulomb
interaction.  The above factors  intricately involved in the
MIT of correlated systems should be seriously considered, 
since a variety of phenomena including
high-$T_\textrm{c}$ superconductivity and colossal magnetoresistance
appear near the filling-controlled MIT.


Recently, filling control over a very wide doping range encompassing
both electron and hole doping has been realized in a two-dimensional
``misfit-layer" compound system La$_{1.17-x}$A$_x$VS$_{3.17}$ (A = Sr
or Pb).\cite{misfit1,misfit2,misfit3,misfit4} The crystal structure is
composed of alternately stacking LaS and VS$_2$ layers.\cite{crystal,
Cava} While the structure of the LaS layer is a distorted rock-salt
type, the VS$_2$ layer forms a two-dimensional triangular lattice of
the V atoms which are octahedrally surrounded by S atoms.  Therefore,
while the lattice constants of LaS and VS$_2$ layers may be
commensurate in one direction, they are inevitably incommensurate in
the other direction.  Remarkably, both electron doping ($x \lesssim
0.16$) and hole doping ($x \gtrsim 0.16$) are possible within the
single system, and therefore La$_{1.17-x}$A$_x$VS$_{3.17}$ is an
unique system for studying the filling-controlled MIT. According to
the electrical resistivity measurement,\cite{misfit1,misfit2,misfit3}
the system become most insulating when the substitution $x$ approaches
$x \simeq 0.16$, where the triply degenerated $t_{2g}$ orbitals of V
3$d$ are filled by two electrons.  Assuming that the doped hole
concentration $\delta$ is given by $\delta \equiv x-0.16$ (and hence
that the doped electron concentration is given by $-\delta\equiv 0.16-x$), the holes
and electrons are doped up to $\delta \simeq 0.19$ and $-\delta \simeq
0.16$, respectively.  Also the Hall coefficient and thermoelectric
power measurements\cite{misfit1,misfit2} have indicated that the
carrier density decreases towards the insulating limit $x\simeq 0.16$
and the carrier changes its sign between $x<0.15$ and $x \geq 0.17$. 
Furthermore, those two transport properties are considerably
temperature dependent in a similar way to the high-$T_\textrm{c}$
cuprates.  Another similarity to the cuprates is that the
low-temperature electronic specific-heat coefficient $\gamma$
\textit{diminishes} towards the Mott insulating state.\cite{misfit3} In the
insulating phase of La$_{1.17-x}$A$_x$VS$_{3.17}$, a nonmagnetic
behavior has been indicated by a nuclear-magnetic-resonance (NMR)
study.\cite{misfit4}  Lattice distortions in the insulating sample $x=0.17$ 
are suggested by anomalies at $\sim 280$ K
in the temperature dependence of the thermal dilatation and the
ultrasound velocity.\cite{misfit3}


In the present study, the electronic structure of
La$_{1.17-x}$Pb$_x$VS$_{3.17}$ is systematically studied by means of
photoemission spectroscopy.  We have investigated the
chemical-potential shift with carrier-doping, and the doping and
temperature dependence of the density of states (DOS) of the V 3$d$
band.  We also discuss possible effects of structural disorder under
strong electron correlations near the MIT.

\section{Experiment}

Polycrystalline samples of La$_{1.17-x}$Pb$_x$VS$_{3.17}$ ($x=0, 0.1,
0.15, 0.17, 0.25, 0.3$, and 0.35) were prepared by sintering a mixture
of La$_2$S$_3$, Pb, V, and S powders at 1050 $-$ 1150 $^\circ$C. X-ray
powder-diffraction patterns indicated that single-phase samples were
obtained.\cite{misfit1} Details of the sample preparation are
described elsewhere.\cite{misfit1,misfit2,misfit3} High-resolution
photoemission spectroscopy was performed for the valence band, using a
GAMMADATA-SCIENTA SES-2002 hemispherical electron-energy analyzer and
a GAMMADATA-SCIENTA VUV-5010 helium discharge lamp.  The overall
instrumental resolution was 4 meV including the natural width
of the excitation source.  The base pressure in the spectrometer
chamber was about $3 \times 10^{-10}$ Torr.  Clean surfaces were
obtained by scraping \textit{in situ} with a diamond file. All the
spectra were recorded within 90 and 60 min after scraping at $T=14$
K and $T=300$ K, respectively, so that the signals of broad spectral
features around $-6$ eV and $-8$ eV due to surface contamination
became inappreciable.  Also x-ray photoemission spectroscopy (XPS)
and x-ray absorption spectroscopy (XAS) were performed for core
levels.  The XPS spectra were collected using a double-pass
cylindrical-mirror analyzer, using the Mg K $\alpha$ line
($h\nu=1253.6$ eV).  The XAS spectra were collected in the total
electron yield method using synchrotron radiation at beamline 2 of
Photon Factory, High Energy Accelerators Research Organization.  The
base pressure in the vacuum chamber was in low $10^{-10} $ Torr range,
and the spectra were taken at liquid-nitrogen temperature $T =
77$ K and at $T = 50$ K for XPS and XAS, respectively.  The overall
energy resolution was approximately 1.0 eV and 0.7 eV for XPS and XAS,
respectively.  Throughout the photoemission measurements, energies
were carefully calibrated using gold evaporated on the sample surface 
just after each series of measurements.

\section{Results}

\subsection{Chemical potential shift}

\begin{figure}[b]
	\includegraphics[width=68mm]{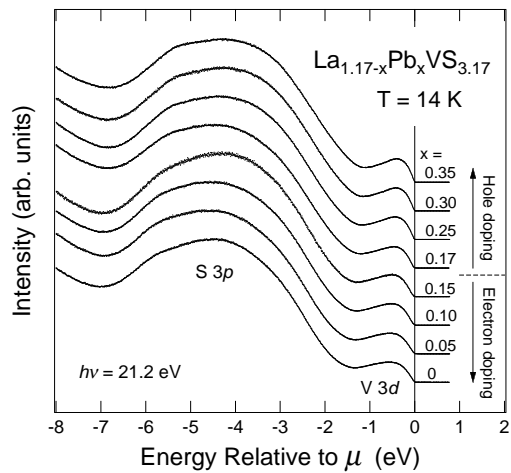}
	\caption{Valence-band spectra of La$_{1.17-x}$Pb$_x$VS$_{3.17}$,
	taken at $T=14$ K using the He \textsc{i} resonance line ($h\nu=$21.2
	eV).  Spectral weight near $\mu$ is depressed for all the
	compositions.}
	\label{valence}
\end{figure}

Valence-band spectra of La$_{1.17-x}$Pb$_x$VS$_{3.17}$ are shown in
Fig.~\ref{valence}.  The broad band around $\sim -4$ eV and the peak
at $\sim -0.5$ eV are mainly derived from the S 3$p$ and V 3$d$
orbitals, respectively.  The spectral weight at $\mu$ is depressed at
$T=14$ K for all the compositions including the metallic samples,
although there should be no energy gap nor a pseudogap in the V 3$d$ band
according to the local-density-approximation (LDA) band-structure
calculation of the parent material VS$_2$.\cite{Myron} 

\begin{figure}
	\includegraphics[width=83mm]{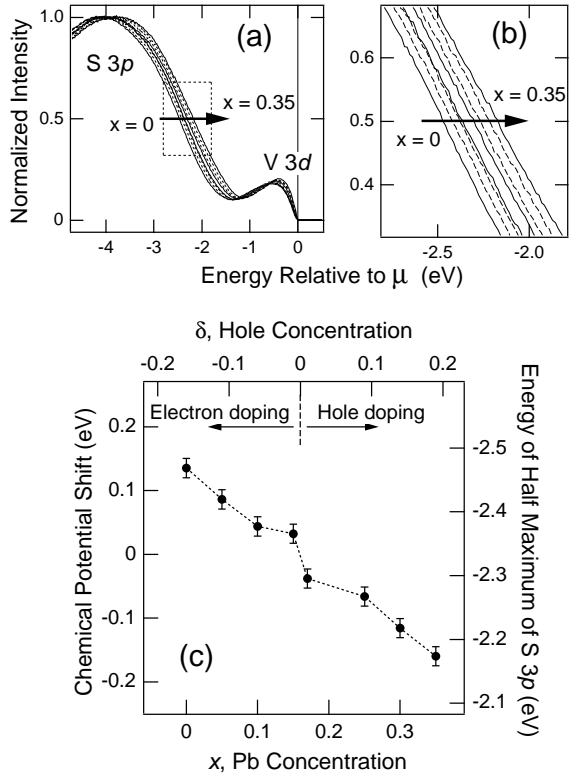}
	\caption{Energy shift of the S $3p$ valence band. 
	(a) Valence-band spectra normalized to the maximum
	spectral intensity around $-4$ eV.  Secondary backgrounds
	have been subtracted from the spectra.  The leading edge of the S 3$p$
	band is rigidly shifted with Pb concentration $x$.  (b) Enlarged
	view of the valence-band spectra around the leading-edge midpoint
	of the S 3$p$ band.  (c) Chemical-potential shift deduced from the
	energy shift of the leading edge of the S 3$p$ band as a function
	of $x$.  The energy shift shows a jump between the electron- and
	hole-doped sides. }
	\label{S3pshift}
\end{figure}

The energy shift of the S 3$p$ band can be accurately determined from
the valence-band spectra, because we find that the spectral shape of the S
3$p$ band remains virtually unchanged except for a rigid energy shift
with Pb substitution $x$.  Although it appears that the intensity of
the secondary background slightly changes with samples in the lower
kinetic-energy region ($E < -5$ eV), the secondary background hardly
affects the spectra in the higher kinetic-energy region ($E > -4$ eV). 
Figures~\ref{S3pshift}(a) and \ref{S3pshift}(b) show that the spectra
of the leading edge of the S $3p$ peak have almost identical curve
in a wide energy range for all the compositions.  
Therefore, we have determined the S 3$p$ shift from the
energy shift of the leading-edge midpoint of the S $3p$ peak,
as shown in Fig.~\ref{S3pshift}(c).

\begin{figure}
	\includegraphics[width=47mm]{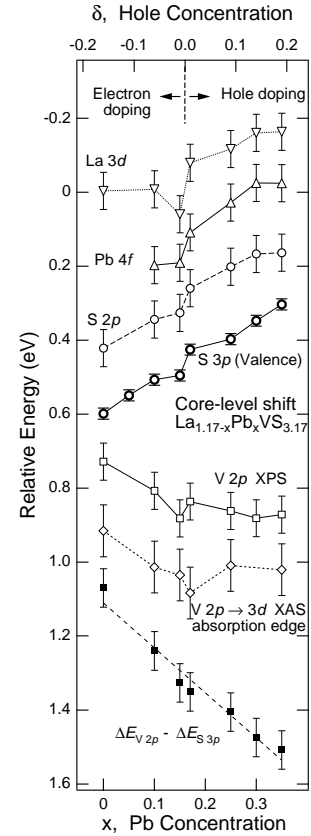}
	\caption{Energy shifts of the La 3$d$, Pb 4$f$, S 2$p$,
	and V 3$d$ core levels.  The energy shift of the V $2p \to 3d$
	x-ray absorption edge is also shown.}
	\label{corelevelshift}
\end{figure}

\begin{figure}
	\includegraphics[width=80mm]{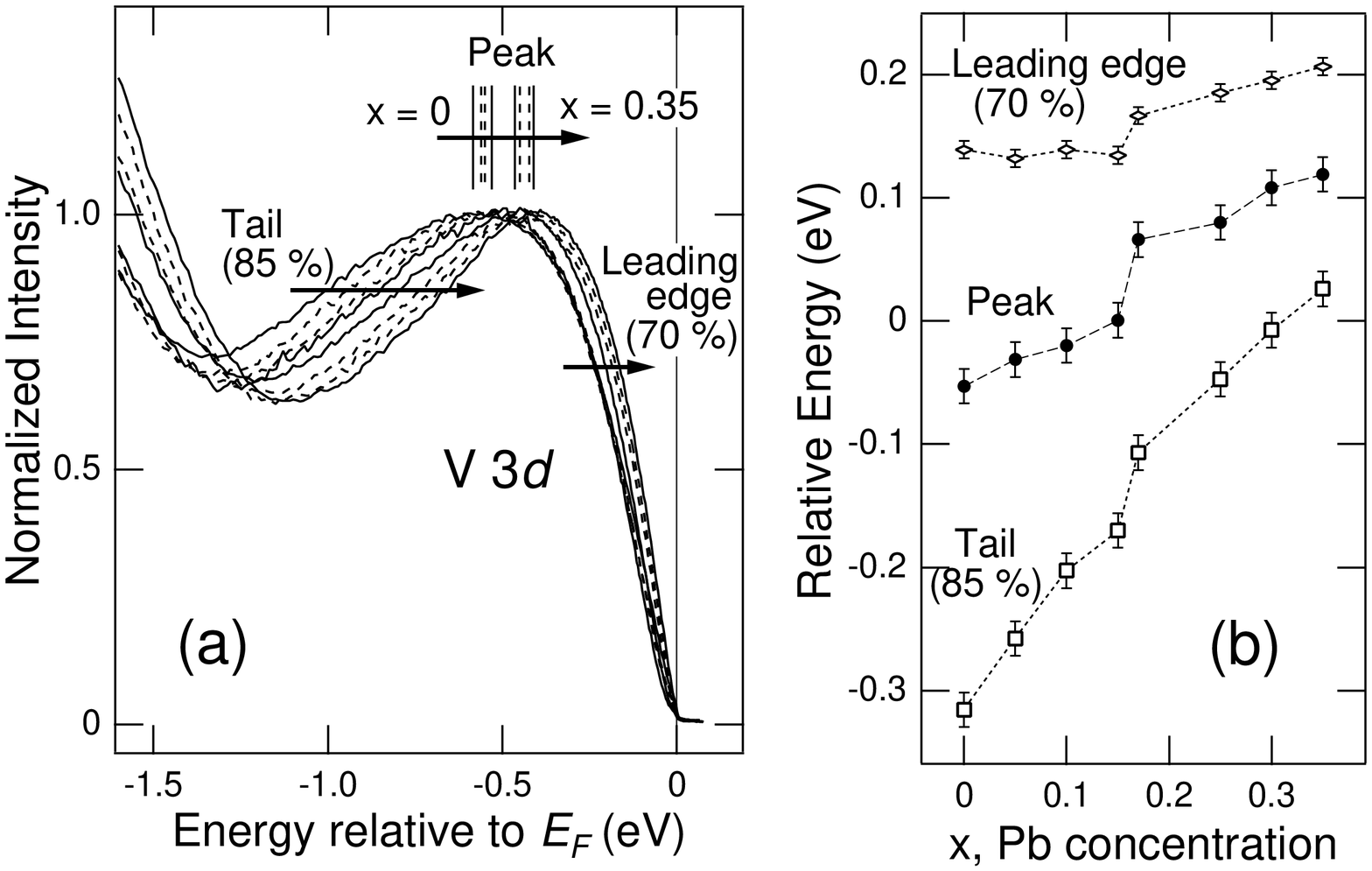}
	\caption{Energy shift of the V $3d$ band. 
	(a) Spectra of the V $3d$ band normalized to the maximum
	spectral intensity at the V $3d$ band peak.  The energy shifts of the
	leading edge and tail of the V $3d$ band are determined, 
	respectively,  by the
	70\% and 85\% level of the V $3d$ maximum intensity around $-0.5$
	eV.  (b) Energy shifts of the leading edge, the peak,
	and the tail of the V $3d$ band.  While the spectral shape of V
	$3d$ band gradually changes with $x$, a discontinuous energy
	shift has also been observed for V $3d$ between $x=0.15$ and 0.17.}
	\label{V3dshift}
\end{figure}


The rigid-band-like shift of the S 3$p$ band is most likely due to the
shift of the chemical potential $\mu$ with carrier doping.  In order
to confirm this, the energy shifts of the core levels have been
measured as in the previous studies on other filling-controlled
materials.\cite{inoshift,harima,satake,kobayashi,JES} As shown in
Fig.~\ref{corelevelshift}, the La, Pb, and S core levels follow the
shift of the S 3$p$ valence band within experimental uncertainties. 
On the other hand, the V core level is shifted in the opposite
direction, i.e., towards higher binding energies with hole doping. 
This shift is caused by the extra contributions
from the varying number of transition-metal $d$ electrons, so-called chemical
shift,  as in the cases of the other filling-controlled
materials.\cite{inoshift,harima,satake,kobayashi,JES} As shown by the
plot at the bottom of Fig.~\ref{corelevelshift}, the energy difference
between V 2$p$ and S 3$p$, $( E_{\text{V } 2p} - E_{\text{S } 3p})$
shows nearly linear dependence on $x$.  Therefore, the observed V 2$p$
shift is decomposed into two components: one is the shift which is
linear in $x$ and the other is the shift which is common to all the
core levels and the valence band.  The former shift is proportional to
the V valence and the latter common shift should reflect the shift of
the chemical potential.  Here, the changes in the Madelung potential
upon substituting Pb for La does not account for the experimental
result, because they should be different between the different atomic
sites, particularly between the cations and anions.  The present
result is remarkably similar to the cases of the other
filling-controlled systems in that all the core levels except for the
core level of the transition-metal element are shifted by the same
amount owing to the chemical-potential
shift.\cite{inoshift,harima,satake,kobayashi,JES}

\begin{figure}
 	\includegraphics[width=56mm]{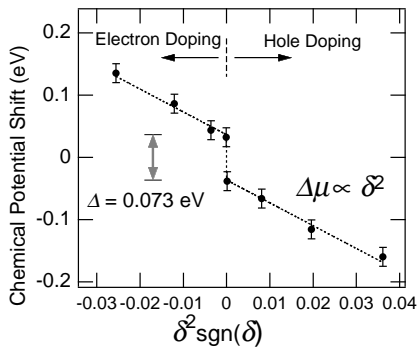}
	\caption{Chemical-potential shift as a function of the square of
	the hole concentration, $\delta^2 \mathrm{sgn}(\delta)$, showing
	a behavior as $\Delta\mu \propto \delta^2$.}
	\label{d2fit}
\end{figure}

Thus, we identify the energy shift of the S 3$p$ valence band shown
in Fig.~\ref{S3pshift}(c) as representing the chemical-potential
shift.  Then, one may find that the chemical potential is abruptly
shifted by $\sim0.07$ eV between $x=0.15$ and 0.17.  It is difficult to
use the V $3d$ band as a direct measure of the chemical-potential
shift, because the spectral shape of the V $3d$  have large
$x$-dependence.  Nevertheless, a discontinuity has also been
observed in the energy shifts of the leading edge, peak and tail of the
V $3d$ band, as shown in Fig.~\ref{V3dshift}.  Therefore, we concluded
that the shift of the chemical potential jumps between $x=0.15$ and
0.17, i.e., across the Mott insulating state.  This suggests that an
energy gap is opened when $x\sim 0.16$ or when the filling of the V
3$d$ band becomes 2 and that the magnitude of the gap is $\sim 0.07$ eV.
When a sufficient number of carriers are doped, namely, in the
overdoped region $|\delta| \gtrsim 0.1$ the chemical potential $\mu$
is monotonously shifted as in a normal metal on both the electron- and
hole-doped sides.  In the underdoped region, on other hand, the shift
of the chemical potential is deviated from the monotonic behavior and
seems to be suppressed towards the insulator limit.  
Since the spectral weight decreases towards $\mu$, 
the observed suppression of the chemical-potential shift is
inconsistent with the rigid band picture
of the conventional band insulator, but in good agreement with
$\Delta\mu \propto \delta^2$, which has been predicted for the
two-dimensional Hubbard model,\cite{Furukawa-Imada}
as shown by dotted lines in Fig.~\ref{d2fit}.

\subsection{Doping dependence of the electronic structure}

\begin{figure}
	\includegraphics[width=84mm]{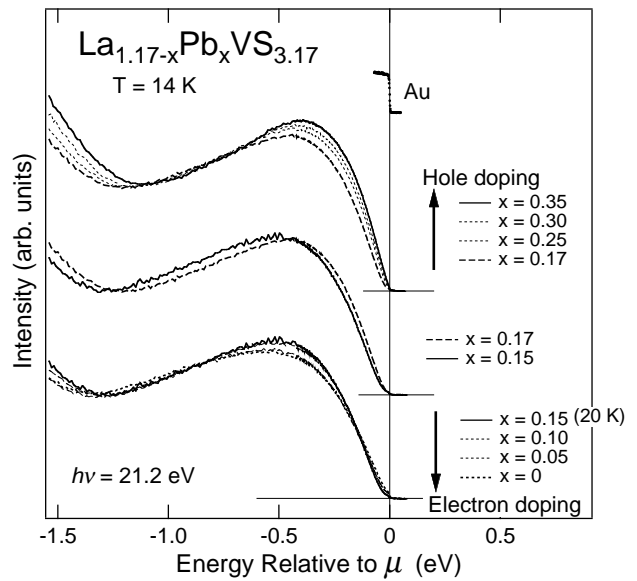}
	\caption{Doping dependence of the V 3$d$ valence-band spectra
	taken at $T=14$ K. The samples $x=0.17$ and 0.15 are slightly
	hole- and electron-doped, respectively, and show a rather large
	difference in spite of the small difference in $x$.}
	\label{DopingV3d}
\end{figure}
\begin{figure}
	\includegraphics[width=88mm]{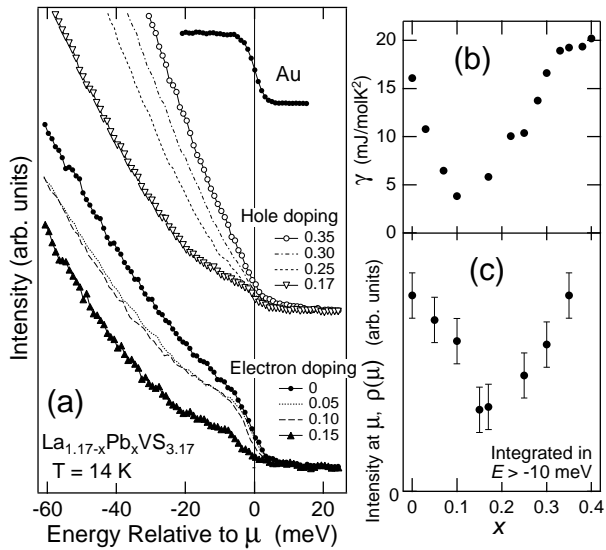}
	\caption{(a) Doping dependence of photoemission spectra around
	$\mu$ taken at $T=14$ K. The spectral intensities for the
	different compositions have been normalized to the maximum
	intensity of the S 3$p$ band.  The gold Fermi edge is also shown. 
	(b) Doping dependence of the low-temperature electronic 
	specific-heat coefficient $\gamma$, taken from
	Ref.~\protect\onlinecite{misfit3}.  (c) Doping dependence of the
	photoemission intensity at $\mu$, $\varrho(\mu)$, approximately
	determined by integrating spectral intensity in the energy range
	of $E>-10$ meV.}
	\label{DopingEF}
\end{figure}

The doping dependence of the V 3$d$ valence-band spectra has been
investigated at low temperature $T=14$ K. The result is shown in
Figs.~\ref{DopingV3d} and \ref{DopingEF}.  The V 3$d$ band shows a
general trend of being shifted towards $\mu$ with substitution of Pb
for La.  Since the spectral intensity at $\mu$ always remains low, the
total width of the V 3$d$ band becomes somewhat narrower
with increasing $x$.  It should be noted that between $x=0.15$ and 0.17, 
the higher binding-energy side of the V 3$d$ band is rigidly shifted
by a large amount in spite of the small change in $x$, 
implying that the chemical potential jumps across the gap
between the electron- and hole-doped samples, as described above.

\begin{figure}
	\includegraphics[width=75mm]{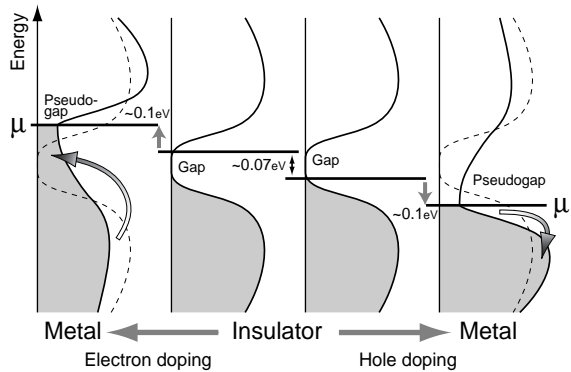}
	\caption{Schematic picture illustrating the evolution of the
	low-energy electronic structure with carrier doping.  The dashed
	curves are the spectrum of the insulator.}
	\label{DrawingDoping}
\end{figure}

Starting from the Mott insulator $x\simeq 0.16$, when the 
electrons are doped, a weak Fermi edge appears as shown in
Fig.~\ref{DopingEF}(a), and the spectral weight within $\sim 0.1$ eV
of $\mu$ is increased at the cost of the decrease in the spectral
weight around $-0.5$ eV. This means that spectral-weight transfer
occurs upon electron doping from $\sim -0.5$ eV to near $\mu$.  If one
takes into account the upward chemical-potential shift by $\sim 0.1$
eV from $x=0.15$ to $x=0$, the spectral-weight transfer into the gap
region should be considered even larger, as schematically illustrated
in  Fig.~\ref{DrawingDoping}.  When the holes are doped into the
$x \simeq 0.16$ Mott insulator, the spectral weight in the vicinity of $\mu$
increases.  However, the increase rate from $x=0.17$ to 0.35 is lower
than that expected from the $x=0.17$ spectrum shifted according to the 
rigid-band model.  This implies that  spectral weight is depleted around
$\mu$ from what would be expected from the rigid-band model 
for all the compositions as a result of pseudogap formation, as
illustrated in Fig.~\ref{DrawingDoping}.

The spectral weight at $\mu$, $\varrho(\mu)$, plotted in
Fig.~\ref{DopingEF} (c) has been 
determined by integrating the spectral intensity in the energy range of
$E > 10$ meV. Although $\varrho(\mu)$ is always small, it shows a
clear doping dependence resembling the doping dependence of
thermodynamic properties, as shown in Fig.~\ref{DopingEF} (c).  Here,
the spectral intensities for the different compositions have been
normalized to the maximum intensity of the S 3$p$ band.  
Corresponding to the fact 
that the electrical resistivity shows the transition to the
insulator in the vicinity of $x\simeq0.16$,\cite{misfit1,misfit2,misfit3}
$\varrho(\mu)$ decreases as $x \to 0.16$, i.e., as the carrier
density decreases either from the electron-doped side or the
hole-doped side.  Figures~\ref{DopingEF}(b) and \ref{DopingEF}(c) show that the
doping dependence of $\varrho(\mu)$ is similar to that of the
low-temperature electronic specific-heat coefficient $\gamma$,\cite{misfit3} 
which is proportional to the QP density at $\mu$, $N^*(\mu)$. 
The doping dependence of $\gamma$ indicates that
no enhancement in the electron effective mass $m^*$ occurs 
near the metal-to-insulator transition in the present system. 
Alternatively, the present photoemission results suggest that 
transition to the insulator is caused by the reduction 
of the effective carrier number due to the pseudogap formation.\cite{Imada}

\subsection{Temperature dependence\\
of the electronic structure}

\begin{figure}
	\includegraphics[width=55mm]{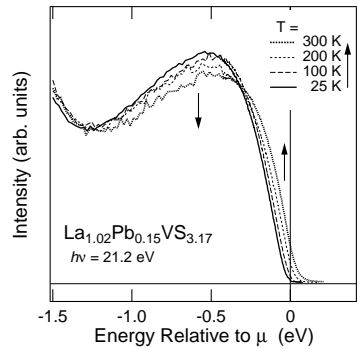}
	\caption{Temperature dependence of the V 3$d$ valence-band
	spectrum for $x=0.15$.  As the temperature increases, the spectral
	weight around $\mu$ increases, while the spectral weight around the
	V 3$d$ peak position is reduced.}
	\label{Tempx15}
\end{figure}

\begin{figure*}
	\includegraphics[width=165mm]{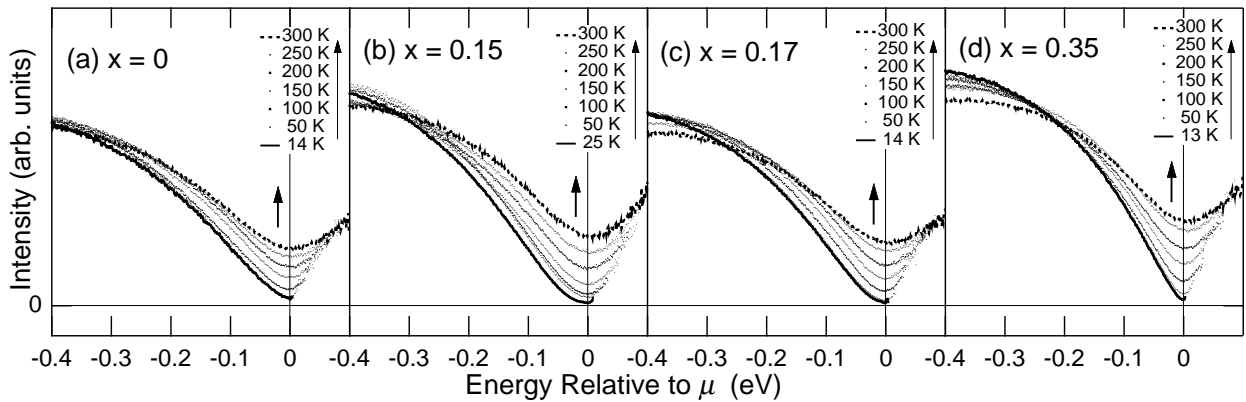}
	\caption{Temperature dependence of the photoemission spectra
	around $\mu$, divided by the Fermi-Dirac distribution function
	for $x=0$, 0.15, 0.17, and 0.35.  It is clearly shown that the
	gap or the pseudogap is rapidly filled with increasing
	temperature.}
	\label{TempAll}
\end{figure*}

\begin{figure}
	\includegraphics[width=44mm]{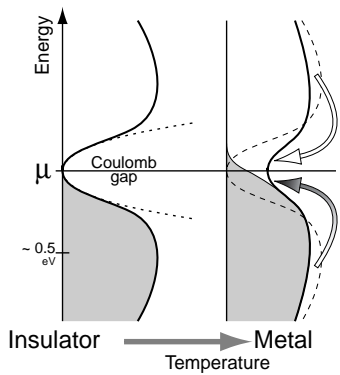}
	\caption{Schematic picture illustrating the evolution of the
	electronic structure with increasing temperature
	for an insulating sample.  }
	\label{DrawingTemp}
\end{figure}

Next, the temperature dependence of the gap and the pseudogap in
La$_{1.17-x}$Pb$_x$VS$_{3.17}$ was investigated. The result for the
insulating sample $x=0.15$ is shown in Fig.~\ref{Tempx15}.  As the
temperature increases, dramatic spectral weight transfer occurs as
shown in the figure: the intensity around $\mu$ ($E \gtrsim -0.3$ eV)
increases, while the peak intensity of the V 3$d$ band ($-1.2 \lesssim E
\lesssim -0.4$ eV) decreases.  Here, the energy scale corresponding to
the temperature change ($4k_BT\lesssim 0.1$ eV) is of the order of the
energy gap $\sim 0.06$ eV, while the spectral weight transfer occurs
over a much wider energy range ($\sim 1$ eV), as have been observed in
many strongly correlated systems.  In order to eliminate the effect of
the temperature dependence of the Fermi-Dirac distribution function
and to extract the intrinsic temperature dependence of the
pseudogap line shape,\cite{Susaki,Greber} the spectra have been divided by the
Fermi-Dirac distribution function, as shown in Fig.~\ref{TempAll}(b). 
The figure clearly shows that the gap of the insulating $x=0.15$
sample is rapidly filled with temperature.  The increase of the
spectral weight at $\mu$ is continuous and monotonic at least from
$T=14$ K to 300 K, and shows no anomaly at $T \simeq 280$ K unlike the
thermal dilatation and the ultrasound velocity. 
Figure~\ref{DrawingTemp} schematically illustrates the evolution of
the electronic structure with increasing temperature.

In Fig.~\ref{TempAll}, temperature-dependent spectra are also shown
for several different composition.  It should be noted that the gap
filling with temperature looks quite similar for all the compositions,
and does not depend on whether it has a real gap or a pseudogap.  At
room temperature 300 K, the spectral intensity at $\mu$ is filled up
to about half of the peak intensity of the V 3$d$ band, so that the
spectra at 300 K appear almost those of a normal metal.

\section{Discussion}


Now, we discuss on the nature of the metal-to-insulator transition in
La$_{1.17-x}$Pb$_x$VS$_{3.17}$.  Since the three $t_{2g}$ orbitals
of the V 3$d$ states are thought to be nearly degenerate,
the insulating state around $x\simeq 0.16$ 
is difficult to be understood as a conventional band insulator,
and should be explained as a Mott insulator induced by
the repulsive electron-electron interaction. In fact, the observed
critical behavior of the chemical-potential shift around MIT
is consistent with the Mott-Hubbard model calculation. 
Then, the nonmagnetic behavior of the insulating state 
of La$_{1.17-x}$Pb$_x$VS$_{3.17}$ at low
temperatures would be interpreted as due to a singlet formation out of
the atomic $S=1$ state of the $d^2$ configuration, e.g.,
the formation of a small cluster such as V$_3$ trimer.\cite{misfit4}
In the case of the weak interaction limit, such a picture
of the insulating state becomes that of a charge-density-wave (CDW) insulator. 
Whether the interaction is strong or weak,
these insulating states require lattice distortion,
which is indeed observed as the anomaly in the thermal dilatation
and in the ultrasound velocity at $T \simeq 280$ K.\cite{misfit3}


Although the chemical-potential shift and the transport properties indicate
that the Mott MIT occurs, no effective mass enhancement has been
observed near the MIT of La$_{1.17-x}$Pb$_x$VS$_{3.17}$ in the
electronic specific-heat coefficient $\gamma$ nor in the
photoemission spectral weight at $\mu$, $\varrho(\mu)$.  Alternatively,
$\varrho(\mu)$ is decreased towards the insulating phase because the
pseudogap develops at $\mu$.  With increasing  temperature,
the pseudogap is rapidly filled.  The behaviors of both 
the chemical-potential shift and the doping and temperature dependence of
$\varrho(\mu)$ are similar to those of the high-$T_\textrm{c}$ cuprate
La$_{2-x}$Sr$_x$CuO$_4$,\cite{ino98,satoLSCO} rather than those of the
three-dimensional typical Mott MIT system
La$_{1-x}$Sr$_x$TiO$_3$.\cite{Yoshida,Kumagai} Therefore, the present
results suggest that the dimensionality of the system is crucial to
whether the mass enhancement occurs or the pseudogap behavior is
observed.  Then, we consider a possible origin of the suppression
of $\varrho(\mu)$ near MIT.  For La$_{2-x}$Sr$_x$CuO$_4$,
the formation of the large pseudogap is closely related to the
antiferromagnetic correlation\cite{ino98} or singlet correlation.\cite{satom} 
Thus, a singlet formation
of the V$_3$ trimer may lead to the disappearance of the electronic
states around $\mu$.  A Hubbard model calculation has shown that
dimerization may reconcile the suppression of the chemical-potential
shift with the absence of the mass enhancement.\cite{Imada}


\begin{figure}
	\includegraphics[width=55mm]{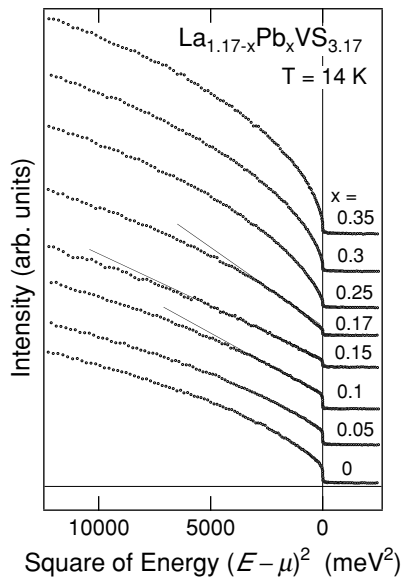}
	\caption{Photoemission spectra near $\mu$ taken at $T=14$ K,
	plotted against the square of binding energy.  The spectrum for
 	$x=0.15$ has the widest straight portion near $\mu$.}
	\label{sqrE}
\end{figure}

In addition, the effect of structural disorder cannot be ignored in the
filling-controlled systems.  In particular, for the present system,
the parent insulator La$_{1.01}$Pb$_{0.16}$VS$_{3.17}$ is
nonstoichiometric unlike the stoichiometric La$_2$CuO$_4$.\cite{mismatch}
(Scraping sample surfaces may also
introduces additional disorder for the present experiment.)  When the
carrier concentration is low,  random potential leads to
Anderson localization.  Doped carriers just being localized are
more strongly affected by the long-range Coulomb interaction between
electrons, and a so-called ``Coulomb gap'' is formed at $\mu$.  Efros
and Shklovskii theoretically predicted that long-range Coulomb interaction in
a disordered insulator drives the electronic states around $\mu$ to
form a soft gap: $N(E)/N_0(\mu) = |(E-\mu)/\Delta_\text{C} |^2$, where
$N_0(\mu)$ is noninteracting DOS at $\mu$.\cite{Efros} Such a Coulomb
gap has also been observed in the photoemission spectra of Fe$_3$O$_4$
and Ti$_4$O$_7$.\cite{Chainani,KobayashiTiO} As seen from
Figs.~\ref{Tempx15} and \ref{TempAll}, the leading edge of the V 3$d$
band is broadened and obscured for all the compositions.  In
particular, for the insulating samples $x=0.15$ and 0.17, the spectral
intensity appears parabolic near $\mu$, $\varrho(E) \propto
(E-\mu)^2$ at low temperatures, as shown in Fig.~\ref{TempAll}.  
In order to see this more clearly, the spectra at low temperatures are plotted as a
function of the square of energy in Fig.~\ref{sqrE}.  While the
spectra for both metallic ends $x=0$ and 0.35 are convex near $\mu$,
the spectrum of the most insulating sample $x = 0.15$ shows a straight
portion over a wide energy range below $\mu$, suggesting the formation
of a Coulomb gap.  A small residual step  at $\mu$ for $x = 0.15$
indicates that the Coulomb-gap opening is incomplete, but it is
difficult at present to judge whether this is an intrinsic feature of
the material or is due to small residual metallic regions in the
sample.  The spectral evolution from the pseudogap to the Coulomb gap
with decreasing carrier concentration is quite similar to that
observed for Fe$_3$O$_4$ and Ti$_4$O$_7$.\cite{Chainani,KobayashiTiO}
In order to estimate the magnitude of $\Delta_\text{C}$, we simply
assume that the noninteracting DOS, $N_0(\mu)$, is given by the DOS
of the band calculation of VS$_2$.\cite{Myron} Then we obtain $\sim
0.15$ eV as an upper limit of $\Delta_\text{C}$.  Therefore,
$\Delta_\text{C}$ appears comparable to the
Mott insulting gap $\sim 0.07$ eV. The observed photoemission spectra
suggest that the long-range Coulomb interaction has impact on the
electronic states around $\mu$, and that the structural disorder is
important in the La$_{1.17-x}$Pb$_x$VS$_{3.17}$ system.

\section{Conclusion}

In conclusion, photoemission experiments have revealed the evolution
of the electronic structure across the filling-controlled MIT in the misfit-layer
La$_{1.17-x}$Pb$_x$VS$_{3.17}$ system.  The jump of the 
chemical-potential between $x=0.15$ and 0.17 indicates the opening of a
charge gap of $\sim 0.07$ eV. The characteristic behavior of the chemical potential
shift is consistent with the picture that the Mott MIT occurs in the present system as
suggested from the transport properties.  When hole or electron carriers are doped
into the Mott insulator, the gap is filled while spectral weight at
$\mu$, $\varrho(\mu)$, is suppressed for all the compositions due to
the pseudogap formation.  With increasing temperature, the gap is rapidly
filled up and $\varrho(\mu)$ increases monotonously.  On approaching
the MIT from the metallic side,  $\varrho(\mu)$ decreases, indicating 
that the effective carrier number vanishes
towards the insulating phase, consistent with the decrease of
the specific-heat coefficient.  
La$_{1.17-x}$Pb$_x$VS$_{3.17}$ is another two-dimensional
filling-controlled MIT system which shows anomalous behaviors similar
to the high-$T_\textrm{c}$ cuprate La$_{2-x}$Sr$_x$CuO$_4$ in the
chemical potential shift and the doping and temperature dependence of
$\varrho(\mu)$.  The present result implies the importance of the
two-dimensionality in the metal-insulator transition.  The observed
Coulomb-gap behavior is attributed to the impact of long-range Coulomb
interaction on the Anderson localized carriers. 
The present result shows that one should
consider the effect of structural disorder, which is
intricately involved to the filling-controlled MIT,
in addition to electron correlation.

\section*{ACKNOWLEDGMENTS}

This work was partly supported by a Grant-in Aid for Research in
Priority Area ``Novel Quantum Phenomena in Transition-Metal Oxides''
from the Ministry of Education, Culture, Sports, Science and
Technology, Japan.  Part of this work has been done under the approval
of the Program Advisory Committee of Photon Factory (PAC-94G361).


\begin{thebibliography}{99}
\bibitem{mismatch}
In addition to the substitution of Pb for La, the mismatch of the
lattice periodicity between the VS$_2$ and LaS layers may induce
effects similar to the ``random potential'' on electrons in the VS$_2$
layer.
\end{thebibliography}
\end{document}